# Nulling interferometry in space does not require a rotating telescope array

François Hénault
Institut de Planétologie et d'Astrophysique de Grenoble
Université Grenoble-Alpes, Centre National de la Recherche Scientifique
B.P. 53, 38041 Grenoble – France

## ABSTRACT

Space borne nulling interferometry in the mid-infrared waveband is one of the most promising techniques for characterizing the atmospheres of extra-solar planets orbiting in the habitable zone of their parent star, and possibly discovering life markers. One of its most difficult challenges is the control of free-flying telescope spacecrafts moving around a central combiner in order to modulating the planet signal, within accuracy better than one micrometer at least. Moreover, the whole array must be reconfigured regularly in order to observe different celestial targets, thus increasing the risk of loosing one or more spacecrafts and aborting the mission before its normal end. In this paper is described a simplified optical configuration where the telescopes do not need to be rotated, and the number of necessary array reconfigurations is minimized. It allows efficient modulation of the planet signal, only making use of rotating prisms or mirrors located into the central combiner. In this paper the general principle of a nulling interferometer with a fixed telescope array is explained. Mathematical relations are established in order to determining the planet modulation signal. Numerical simulations are carried out for three different arrangements of the collecting telescopes. They confirm that nulling interferometry in space does not require a rotating telescope array.

## 1 INTRODUCTION

Nulling interferometry in the mid-infrared waveband from a space observatory is one of the most promising techniques for characterizing the atmospheres of extra-solar planets orbiting in the habitable zone of their parent star, and possibly discovering gaseous life markers. The question of probing extra-terrestrial life has fundamental scientific and philosophical implications; and has been the subject of numerous technical feasibility studies for more than twenty years. It all started with the reference article from Bracewell, suggesting that a two-telescope, "nulling" interferometer staring at distant stars would make this discovery possible [1]. In Europe, it gave birth to the Darwin project managed by the European Space Agency [2-4]. In the mean time, the National Aeronautics and Space Administration (NASA) of America undertook the study of a Terrestrial Planet Finder Interferometer (TPF-I) with similar objectives [5]. Later, simplified designs aiming at characterizing the atmospheres of giant gaseous extrasolar planets were proposed, such as Pegase [6] and the Fourier Kelvin Stellar Interferometer [7]. Finally, it should be noted that nulling interferometry from ground telescope arrays is still envisioned [8].

The general scheme of a nulling interferometer is illustrated in Figure 1. It consists in an array of collecting telescopes rotating around a central beam combiner, where a destructive interference is created on the central star by means of achromatic phase-shifters. Thus the on-axis starlight is cancelled and the light emitted by any off-axis extrasolar planet can be sensed and analysed spectrally. However this planet signal must be modulated temporally in order to discriminating it from the background. This is achieved by rotating the whole array around the main optical axis by an angle $\theta$ (blue arrows in Figure 1-1). Planet signals are later processed by applying different numerical algorithms that are described in a series of reference papers [9-12].

One realizes that the most difficulty of this technique is to controlling the motions of all the free-flying telescopes and combiner, within accuracy better than one micrometer typically. Moreover, the whole array must be reconfigured regularly in order to point at different celestial target located at angular coordinates ($\alpha$, $\beta$) with respect to the initial configuration. The required movements are indicated by grey arrows in Figure 1-2. It follows that the sum of all the required degrees of freedom consumes a significant amount of fuel, and will limit the duration of the space mission. Moreover, they increase the risks of loosing one or more spacecrafts and aborting the mission before its normal end. These are the reasons why a concept avoiding the need of a rotating telescopes array would be of high interest in future nulling interferometry space missions. In this paper is described a simplified optical configuration with fixed telescopes only, and minimizing the number of necessary reconfigurations. The general description of the design is provided in section 2. Mathematical relations determining the planet modulation signal are given in section 3. Numerical simulations of the planet signal for three different arrangements of the collecting telescopes are presented in section 4 and compared to the cases of rotating arrays. Finally, brief conclusions of the study are drawn in section 5.

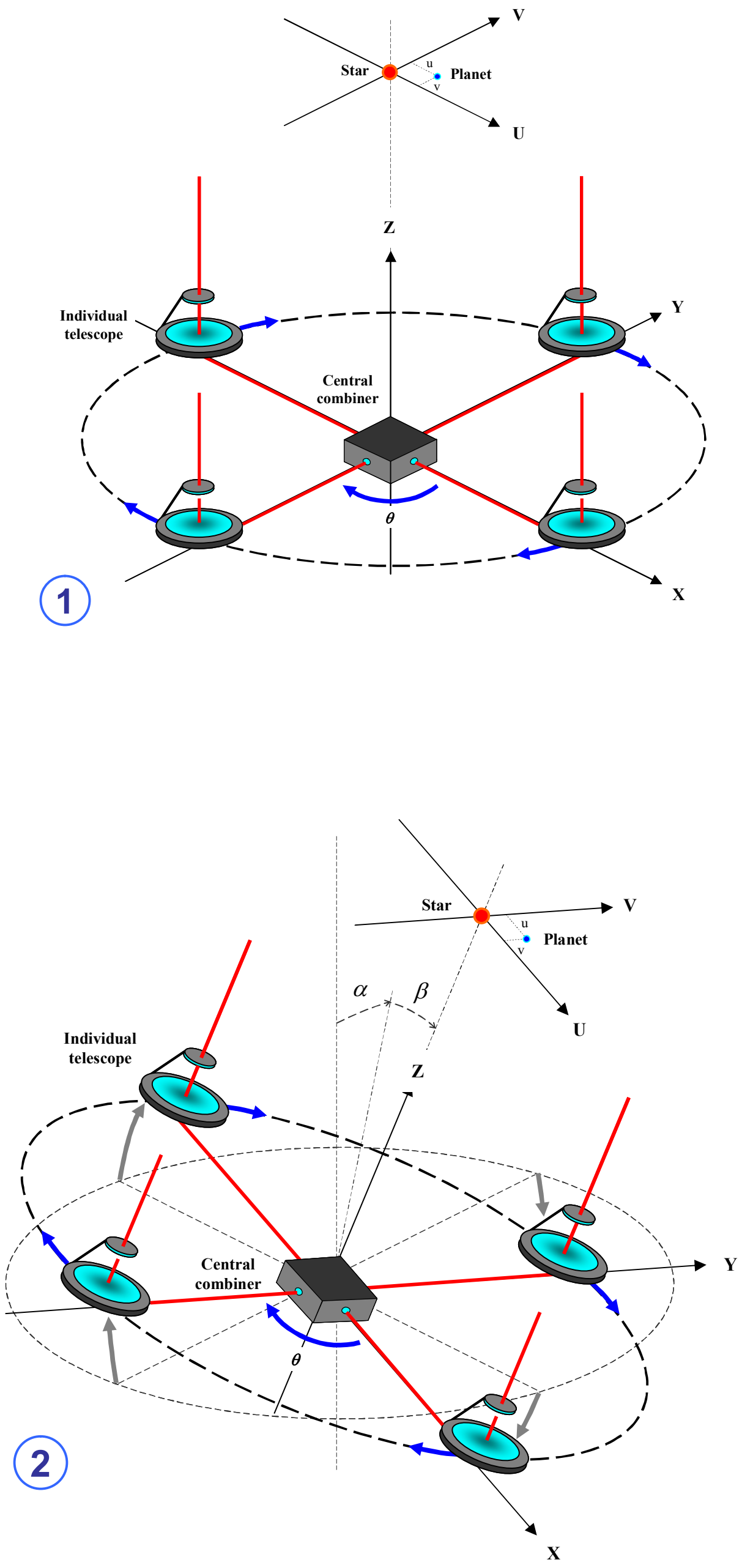

Figure 1: Sketch of a nulling interferometer with rotating telescopes. 1) During observation with blue arrows indicating the rotation motions. 2) Needed array reconfiguration motions to pointing at different sky targets (grey arrows).

# 2 CONCEPTUAL DESIGN

In this section is firstly explained the principle of planet signal modulation without rotating the telescope array (subsection 2.1). Then brief descriptions of the array configuration and delay lines systems are given in subsection 2.2.

## 2.1 Modulating the planet signal

The basic principle for planet signal modulation with a fixed interferometer array is illustrated in Figure 2. It shows two telescopes staring at the central star from the entrance pupil plane OXY, whose beams are combined multi-axially onto Single-mode waveguides (SMW) located at the final focal plane O"X"Y" of the interferometer, where they filter the Wavefront errors (WFE) of the telescopes and all preceding optics. It is assumed that achromatic phase-shifters (not shown in the Figure) provide a high extinction ratio to the central star and maximal contrast to the off-axis extra-solar planet. Planet modulation is introduced by a pupil rotator system located just before the multi-axial combining optics: it can either be made of a Dove prism or of a K-mirror rotating by an angle $\theta$' around the main optical axis Z. The pupil rotator, phase-shifters, focusing optics and SMW are all located inside the central combiner spacecraft. Thus planet modulation is achieved by virtue of the rotating exit sub-pupils in the combining optics plane O'X'Y' only. The mathematical justification of such assertion will be the scope of section 3.

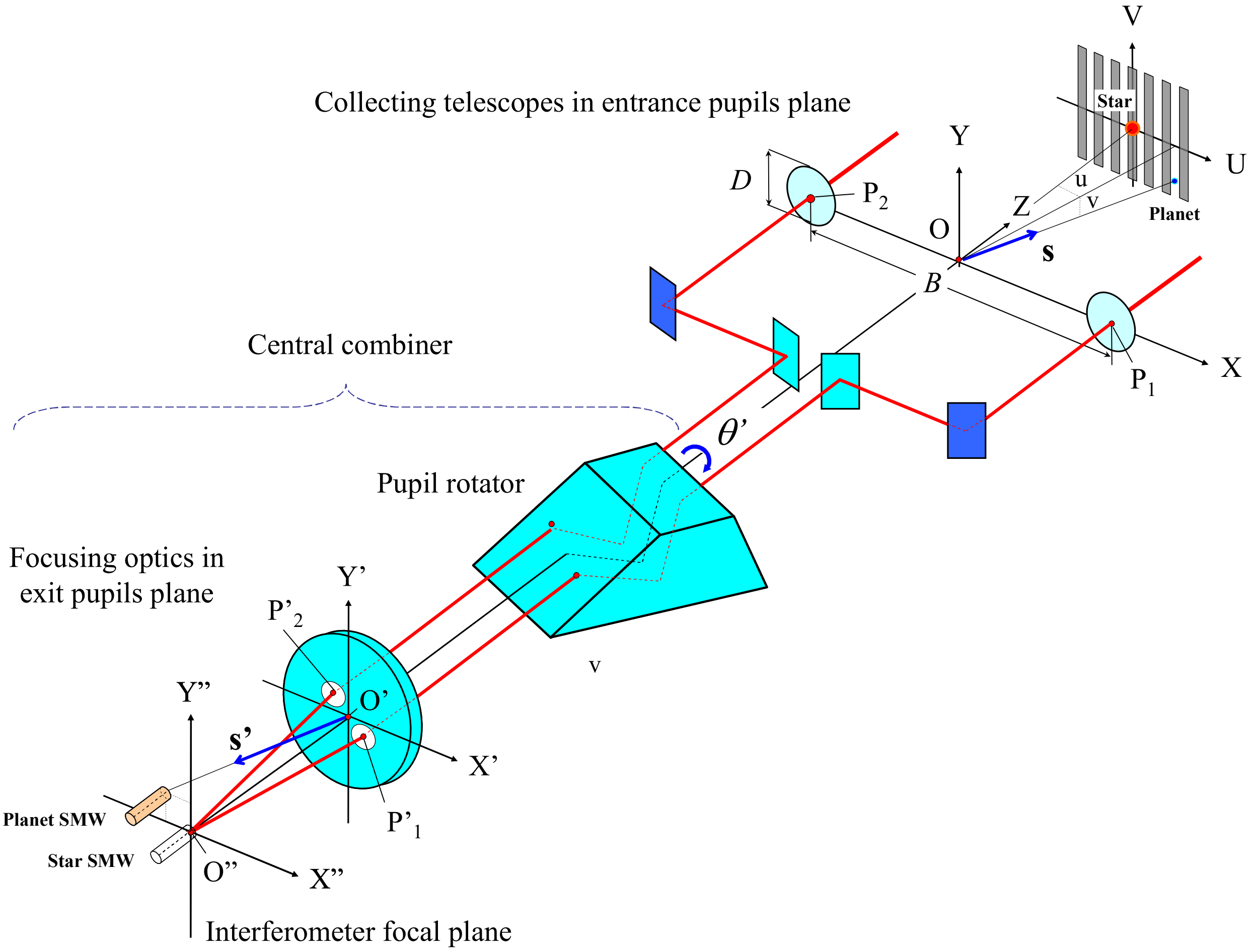

Figure 2: Principle of signals modulation inside the central beam combiner. It is ensured by a pupil rotator located behind the final focusing optics. Two Single-mode waveguides (SMW) are filtering the wavefronts in the interferometer focal plane, one for the central star and the other for the extrasolar planet. The central SMW is only used for monitoring purpose.

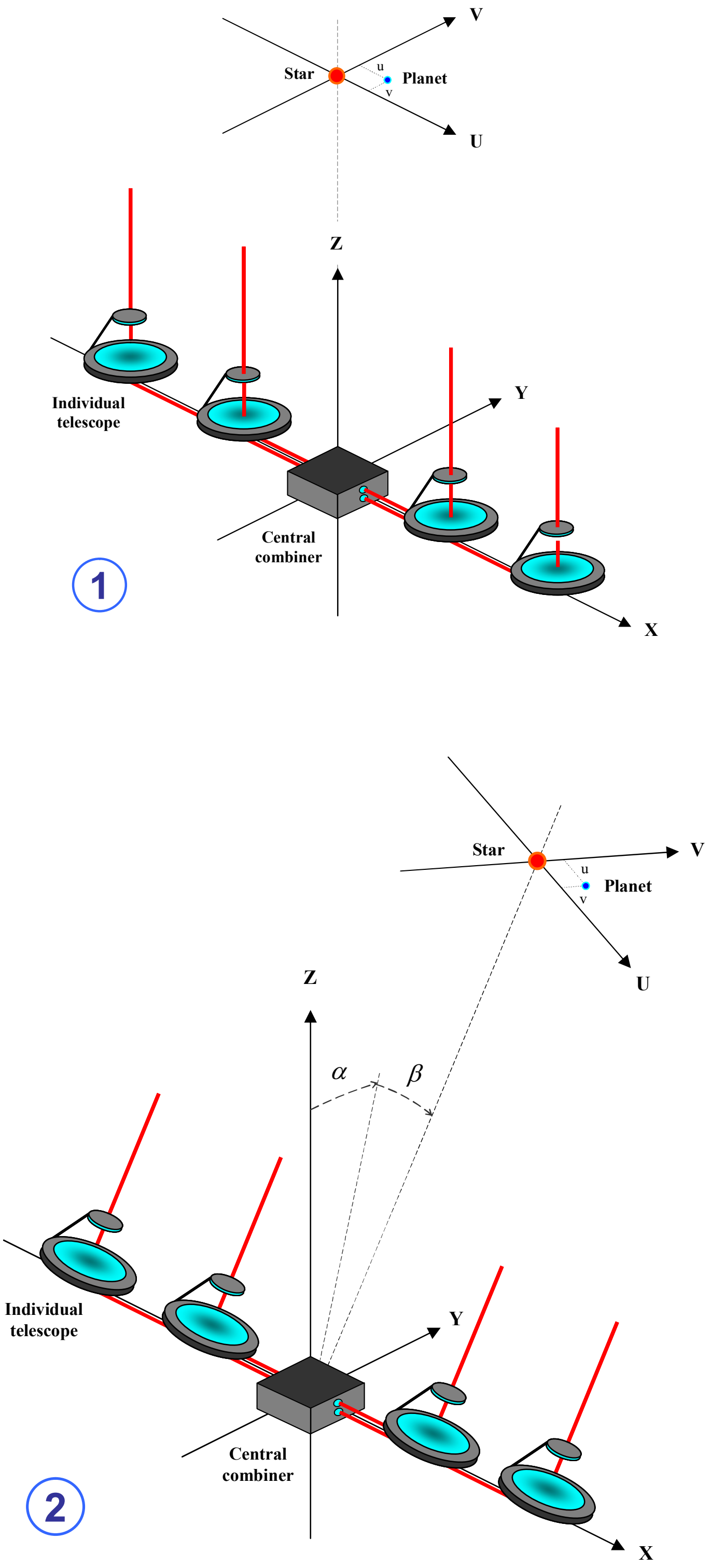

Figure 3: Linear interferometer with a fixed telescope array. 1) On-axis observation. 2) Off-axis observation with no needed array reconfiguration.

### 2.2 Interferometer configuration and delay line systems

The generalization of the previous design to an array comprising four or more telescopes is illustrated in Figure 3. Here all telescopes and the central combiner are aligned along the single X-axis, thus forming a linear array interferometer. Figure 3-1 shows a starting configuration where the direction of the observed star is parallel to the main optical axis Z. No rotation around this axis is needed. All collecting telescopes are assumed to be mounted on altazimuthal drives, allowing them to point at other sky targets located at other angular coordinates $\alpha$ and $\beta$ (see Figure 3-2). Here again no reconfiguration movements of the array is required, as long as the generated Optical path differences (OPD) with respect to the combiner do not exceed the range of the delay lines onboard of the telescopes and combiner spacecrafts. From the theoretical relations given in subsection 3.2 the OPD to be compensated for should be equal to $B \cos\alpha \cos\beta / 2$; with $B$ the entrance baseline between two adjacent telescopes. Assuming that the observed stellar systems stay inside a cone of 15 deg. angle with respect to the Z-axis and that double pass delay lines can be shared between the collecting telescopes and central combiner spacecrafts leads to a maximal OPD range equal to $B\left(1+\sqrt{3}\right)/8\sqrt{2}$ = 0.43 m for a baseline $B$ = 100 m. Designs of such long range delay lines already exist [13-15] and could be integrated within the volumes allocated for each spacecraft. A simplified scheme of such delay line systems is depicted in Figure 4 for the case of a 4-telescope linear interferometer array. Therefore only periodical reorientation of the array would be needed to achieving full sky coverage.

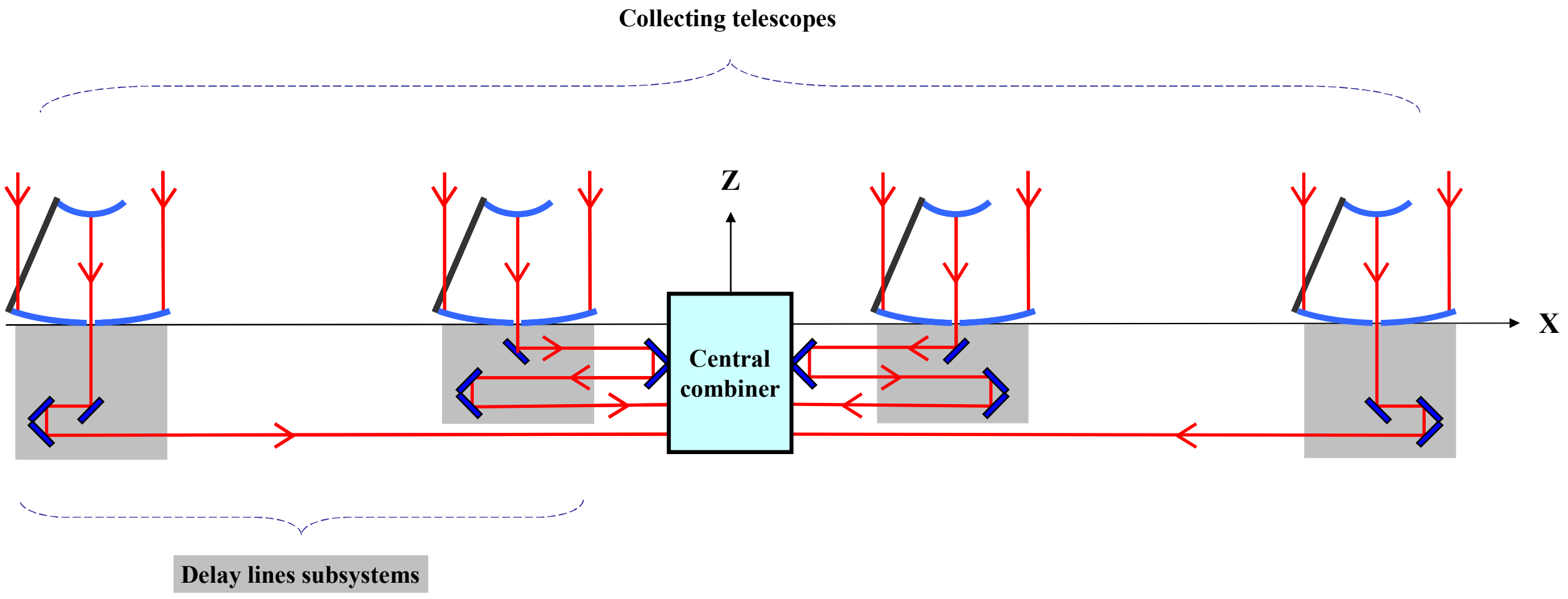

Figure 4: Conceptual design of delay line systems for a 4-telescope linear interferometer array.

## 3 THEORETICAL ANALYSIS

### 3.1 Coordinate frames and scientific notations

As illustrated in Figure 1 to Figure 3, the employed coordinate systems are defined as:

- An entrance pupil reference frame (OXYZ) where OZ is the main optical axis, and all entrance sub-pupils of the collecting telescopes are located in the same OXY plane. The pupil centres are defined by their Cartesian coordinates (x,y). In this reference frame the direction of the extrasolar planet is defined by a unitary vector **s** = (sin u cos v; sin v; cos u cos v) ≈ (u; v; 1) owing to the very small values of tits angular coordinates u and v.
- An exit pupil reference frame (O'X'Y'Z) where the O'X'Y' plane is located at the output pupil of the beam combiner. The centres of the exit sub-pupils are defined by their Cartesian coordinates (x',y') in that plane. In this reference frame a unitary vector **s'** is directed to the planet SMW located in the final detection plane (see

Figure 2). The directing cosines of this vector can be approximated as (–u'; –v'; –1) with the same assumption as before.

- A reference frame (O"X"Y"Z) attached to the final detection plane, where SMW locations are defined by their Cartesian coordinates (x",y"). These coordinates are linked to the vector **s'** by the relations x"= *F*' u' and y"= *F*' v' with *F*' the focal length of the focusing optics inside the combiner. Thus they are redundant notations and are unused in the following study.

Let us now assume that the optical conjugations between the entrance telescope pupils in the OXY plane and the exit sub-pupils in plane O'X'Y' are perfectly stigmatic. Then the following scientific notations are used (bold characters denoting vectors):

| | |
|---|---|
| $i$ | Complex square root of –1 |
| $k$ | Wavenumber $2\pi/\lambda$ of the electro-magnetic field assumed to be monochromatic, and $\lambda$ is its wavelength |
| $\Omega$, $d\Omega$ | The full observed Field of View (FoV) in terms of solid angle and its integration element |
| $O(\mathbf{s})$ | Angular brightness distribution of the observed sky scene, here an extrasolar planet system |
| N | Number of collecting telescopes in the array |
| $a_n$ | Amplitude transmission factor from the n<sup>th</sup> telescope ($1 \leq n \leq N$) |
| $\varphi_n$ | Phase-shift added to the n<sup>th</sup> telescope beam inside the beam combiner ($1 \leq n \leq N$), by means of achromatic phase-shifters |
| $B$ | Distance between two adjacent telescopes in the OXY plane |
| $\theta$ | Rotation angle of the telescope array around Z-axis |
| $D$ | Diameters of the collecting telescopes, assumed to be identical for all of them |
| $PSF_T(\mathbf{s'})$ | The Point Spread Function (PSF) of an individual collecting telescope. For aberration-free optics it is equal to the classical Airy distribution $|2J_1(\rho)/\rho|^2$, where $\rho = k\,D\,\|\mathbf{s'}\| / 2$ in angular coordinates, and $J_1$ is the type-J Bessel function at the first order |
| $\mathbf{OP_n}$ | Vector defining the entrance pupil centre $P_n$ of the n<sup>th</sup> telescope in the OXY plane ($1 \leq n \leq N$) |
| $B'$ | Distance between two adjacent exit sub-pupils in the O'X'Y' plane |
| $\theta'$ | Pupil rotator angle around Z-axis |
| $D'$ | Diameters of the exit sub-pupils in the O'X'Y' plane, assumed to be identical |
| $\mathbf{O'P'_n}$ | Vector defining the center $P'_n$ of the n<sup>th</sup> sub-pupil in the O'X'Y' plane ($1 \leq n \leq N$) |
| $d$ | "Dilution factor" of the exit sub-pupils, equal to $d = D\,B' / D'\,B$. It typically ranges from zero (case of an axial combiner) to unity (case of a Fizeau imaging interferometer with multi-axial combining optics) |

## 3.2 General relations

Let us make use of a mathematical formalism defined in previous publications where the intensity distribution in the O"X"Y" plane is retro-projected onto the sky, so that it is independent of the combining optics and detector parameters. Mathematical justifications of this formalism can be found in Refs. [16-17]. Then the expression of the image $I(\mathbf{s'})$ formed by the nulling interferometer can be written as:

$$I(\mathbf{s}') = \iint_{\mathbf{s} \in \Omega} O(\mathbf{s})\; PSF_T(\mathbf{s}' - \mathbf{s})\; F(\mathbf{s},\mathbf{s}')\, d\Omega, \tag{1a}$$

where the function $F(\mathbf{s},\mathbf{s}')$ is defined as:

$$F(\mathbf{s},\mathbf{s}') = \left| \sum_{n=1}^{N} a_n \exp[i\varphi_n] \exp[ik\,\xi_n(\mathbf{s},\mathbf{s}')] \right|^2, \tag{1b}$$

and $\xi_n(\mathbf{s},\mathbf{s}')$ stands for an extra OPD term resulting from the angular deviations of vectors **s** and **s'** with respect to the Z-axis, here equal to:

$$\xi_n(\mathbf{s},\mathbf{s}') = \mathbf{s}\,\mathbf{OP_n} - d\;B\;\mathbf{s}'\mathbf{O'P'_n}\;/B', \tag{2}$$

using the mathematical notations defined in the previous subsection.

Let us firstly consider the case of the signal emitted by the central star and incoming into the planet SMW located in the O"X"Y" plane (see f Figure 2). Here the observed sky scene O(**s**) can be approximated to a Dirac delta function $\delta(\mathbf{s})$, hence leading to the simplified expression of I(**s'**), here re-noted S(**s'**):

$$S(\mathbf{s}') = PSF_T(\mathbf{s}')\;F(\mathbf{0},\mathbf{s}') \tag{3}$$

with **0** the null vector. Similarly, the signal recorded by the planet SMW originates from an off-axis Dirac delta function $\delta(\mathbf{s}-\mathbf{s_P})$ in the direction $\mathbf{s_P} \approx (u_P,v_P,1)$ of the extrasolar planet, thus the planet signal P($\mathbf{s_P}$,**s'**) writes:

$$P(\mathbf{s_P},\mathbf{s}') = PSF_T(\mathbf{s}'-\mathbf{s_P})F(\mathbf{s_P},\mathbf{s}') \tag{4}$$

### 3.3 Analytical expression of the OPD for three different pupils configurations

Three different array configurations are considered in this study. The first one is the original Bracewell's two-telescope proposal [1]. The two other involve four collecting telescopes following a linear geometry. They are the Degenerated Angel Cross (DAC) proposed by Angel and Woolf [18-19] and the Dual Chopped Bracewell (DCB) analyzed in Ref. [11]. The geometrical configurations of these arrays are depicted in Figure 5. Their input parameters, including amplitude transmission $a_n$, phase-shifts $\varphi_n$, and the geometry of their input and output sub-pupil are summarized in Table 1, whether the array is rotated or not.

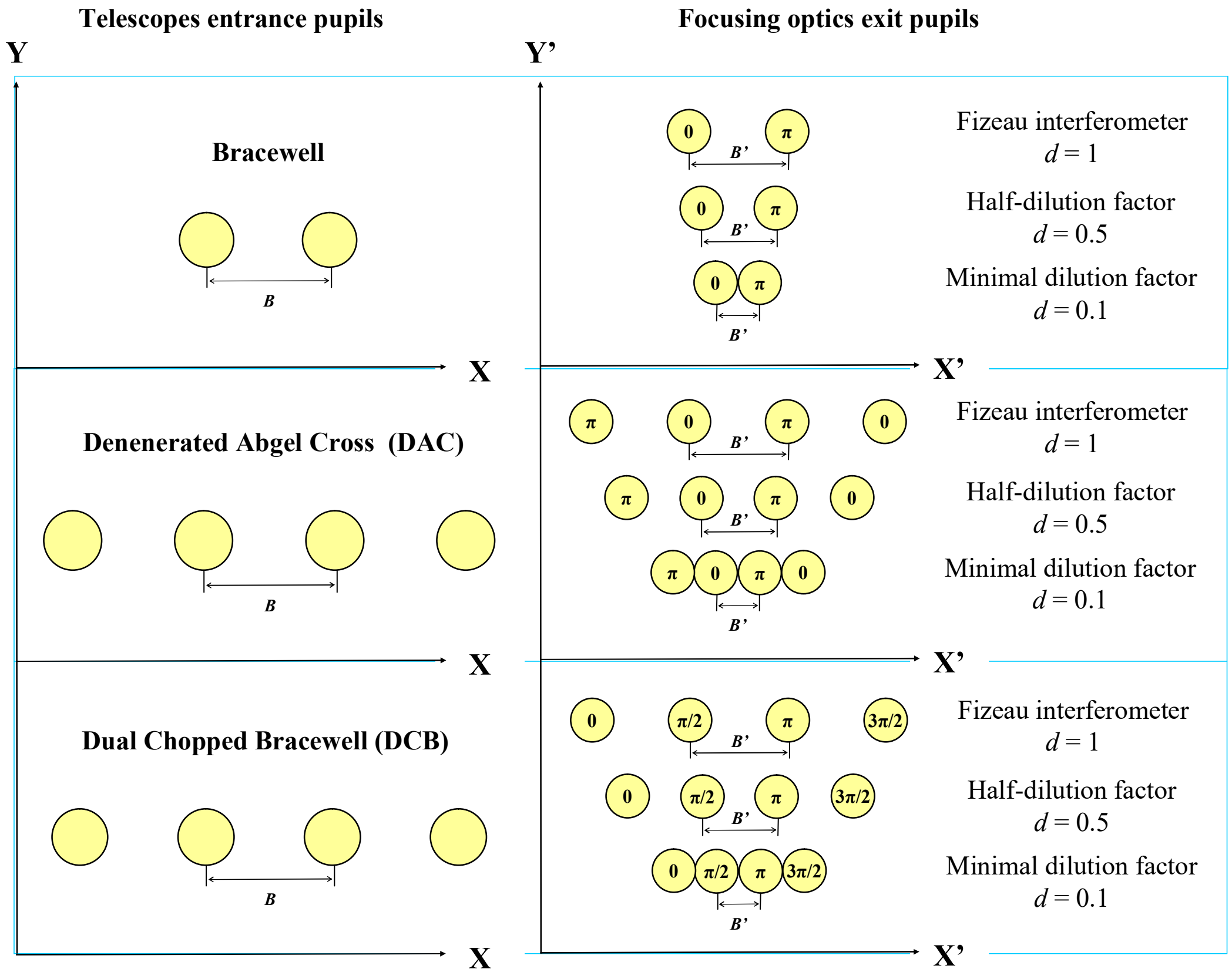

Figure 5: Entrance and exit pupil configurations of the three studied interferometer arrays.

Table 1: Input parameters of the three studied interferometer arrays.

| Parameters | Symbol | Array configuration: Two-telescope Bracewell | Degenerated Angel Cross (DAC) | Linear Dual Chopped Bracewell (DCB) | |
|---|---|---|---|---|---|
| Telescopes number | N | 2 | 4 | 4 | |
| Amplitude transmission | $a_n$ | 0.5<br>0.5 | 0.25<br>0.25<br>0.25<br>0.25 | 0.25<br>0.25<br>0.25<br>0.25 | |
| Phase-shifts | $\varphi_n$ | 0<br>π | 0<br>π<br>0<br>π | 0<br>π/2<br>π<br>3π/2 | |
| Entrance pupils coordinates (non rotating) | $\mathbf{OP_n}$ | $(-B/2, 0)$<br>$(+B/2, 0)$ | $(-3B/2, 0)$<br>$(-B/2, 0)$<br>$(+B/2, 0)$<br>$(+3B/2, 0)$ | $(-3B/2, 0)$<br>$(-B/2, 0)$<br>$(+B/2, 0)$<br>$(+3B/2, 0)$ | (5a) |
| Entrance pupils coordinates (rotating) | $\mathbf{OP_n}$ | $-B/2(\cos\theta, \sin\theta)$<br>$+B/2(\cos\theta, \sin\theta)$ | $-3B/2(\cos\theta, \sin\theta)$<br>$-B/2(\cos\theta, \sin\theta)$<br>$+B/2(\cos\theta, \sin\theta)$<br>$+3B/2(\cos\theta, \sin\theta)$ | $-3B/2(\cos\theta, \sin\theta)$<br>$-B/2(\cos\theta, \sin\theta)$<br>$+B/2(\cos\theta, \sin\theta)$<br>$+3B/2(\cos\theta, \sin\theta)$ | (5b) |
| Exit pupils coordinates (non rotating) | $\mathbf{O'P'_n}$ | $(-B'/2, 0)$<br>$(+B'2, 0)$ | $(-3B'/2, 0)$<br>$(-B'/2, 0)$<br>$(+B'/2, 0)$<br>$(+3B'/2, 0)$ | $(-3B'/2, 0)$<br>$(-B'/2, 0)$<br>$(+B'/2, 0)$<br>$(+3B'/2, 0)$ | (5c) |
| Exit pupils coordinates (rotating) | $\mathbf{O'P'_n}$ | $-B'/2(\cos\theta', \sin\theta')$<br>$+B'/2(\cos\theta', \sin\theta'))$ | $-3B'/2(\cos\theta', \sin\theta')$<br>$-B'/2(\cos\theta', \sin\theta')$<br>$+B'/2(\cos\theta', \sin\theta')$<br>$+3B'/2(\cos\theta', \sin\theta')$ | $-3B'/2(\cos\theta', \sin\theta')$<br>$-B'/2(\cos\theta', \sin\theta')$<br>$+B'/2(\cos\theta', \sin\theta')$<br>$+3B'/2(\cos\theta', \sin\theta')$ | (5d) |

Inserting the expressions of vectors $\mathbf{OP_n}$ and $\mathbf{O'P'_n}$ given in Table 1 into Eq. 2 allows writing the differential OPD $\xi_n(\mathbf{s},\mathbf{s}')$ as:

$$\xi_n(\mathbf{s},\mathbf{s}') = [n - 0.5 - N/2]\,\xi_0(\mathbf{s},\mathbf{s}') \qquad \text{with N = 2 or 4,} \tag{6}$$

and $\xi_0(\mathbf{s},\mathbf{s}')$ is a generic expression depending on the array rotation movements, i.e.

1) For a rotating telescope array with $\theta = \theta' \neq 0$

$$\xi_0(\mathbf{s},\mathbf{s}') = B(\mathrm{u}\cos\theta + \mathrm{v}\sin\theta) - \frac{DB'}{D'}(\mathrm{u}'\cos\theta + \mathrm{v}'\sin\theta) = B\,\mathbf{r}(\mathbf{s} - d\,\mathbf{s}'), \tag{7}$$

where $\mathbf{r}$ is the rotation vector equal to $(\cos\theta, \sin\theta)$.

2) For the fixed telescope array with $\theta = 0$ and $\theta' \neq 0$

$$\xi_0(\mathbf{s},\mathbf{s}') = B\,\mathrm{u} - \frac{DB'}{D'}(\mathrm{u}'\cos\theta' + \mathrm{v}'\sin\theta') = B(\mathbf{i}\,\mathbf{s} - d\,\mathbf{r}'\mathbf{s}'), \tag{8}$$

with rotation vector **r'** equal to (cos$\theta$', sin$\theta$'), and **i** is the unitary vector along the X-axis of the entrance pupils plane. Here it must be noted that the angular coordinate v of the extrasolar planet disappears from the expression of $\xi_0(\mathbf{s},\mathbf{s}')$. Hence an indetermination occurs on its precise location along the V-axis (see Figure 3). It follows that the main advantage provided by the non-rotating telescopes concept is somewhat counterbalanced by such loss of information, although the u coordinate of the planet remains accessible. Also, Eq. 8 evidences the crucial role played by the dilution factor parameter $d$ : when the interferometer makes use of axial beam combining (e.g. by means of beamsplitters) $d$ tends toward zero and the modulation term vanishes. Thus only multi-axial combiners for which $d \neq 0$ will be considered in the remainder of the study.

Inserting now Eq. 6 into Eq. 1b and expanding the square modulus, the function $\mathrm{F}(\mathbf{s},\mathbf{s}')$ finally writes for the three considered array configurations as:

Bracewell interferometer $$\mathrm{F}(\mathbf{s},\mathbf{s}') = (1 - \cos[k\xi_0(\mathbf{s},\mathbf{s}')])/2 = \sin^2[k\xi_0(\mathbf{s},\mathbf{s}')] \tag{9}$$

Degenerated Angel Cross $$\mathrm{F}(\mathbf{s},\mathbf{s}') = (4 - 6\cos[k\xi_0(\mathbf{s},\mathbf{s}')] + 4\cos[2k\xi_0(\mathbf{s},\mathbf{s}')] - 2\cos[3k\xi_0(\mathbf{s},\mathbf{s}')])/16 \tag{10}$$

Linear Dual Chopped Bracewell $$\mathrm{F}(\mathbf{s},\mathbf{s}') = (4 - 6\sin[k\xi_0(\mathbf{s},\mathbf{s}')] - 4\cos[2k\xi_0(\mathbf{s},\mathbf{s}')] + 2\sin[3k\xi_0(\mathbf{s},\mathbf{s}')])/16 \tag{11}$$

Numerical simulations based on relations 3–4, 7–8 and 9–11 are presented in the next section for both the cases of a rotating and fixed telescope array.

# 4 NUMERICAL SIMULATIONS

## 4.1 Assumptions – Framework of the study

As mentioned in the introduction, the main goal of this study is to trade both the non-rotating vs. rotating telescope array designs, in light of the achieved planet modulation signals. Thus the numerical model used in this section is only based on the mathematical relations presented in section 3, regardless of known bias and uncertainty sources well documented in the existing scientific literature. To name just a few, the reader is invited to refer to [10] and [20] for theoretical aspects, and to [21-22] for experimental ones. Moreover, it is expected that the performance of the interferometers should be affected in the same way for both the fixed and rotating telescopes options. Hence the following assumptions are made:

1) All telescopes are aberration-free, as well as their downstream optics,
2) Entrance and exit sub-pupils optical l conjugation is perfectly stigmatic,
3) There are no telescope pointing errors, and both the pupil rotator and its driving mechanism are aligned ideally with respect to the Z-axis,
4) Any achromatic phase-shifter or wavefront error filtering devices can be employed without loss of generality,
5) Only the pupil rotator modulates the signal of the planet. No additional internal modulation schemes such as described in Refs. [11] and [23] are introduced,
6) No detailed noise budget is introduced. Instead, a maximal contrast value of $10^6$ is assumed for the star and planet signals, including shot and detector noises.
7) The central star is unresolved, thus no stellar leakage is present,
8) Local and extrasolar zodiacal light backgrounds are neglected.

Under such hypotheses, numerical results for both the cases of rotating and fixed telescopes arrays are presented in the next sub-section.

### 4.2 Numerical results

Numerical simulations are carried out for the nine cases depicted in Figure 5, namely the original Bracewell, DAC and DCB configurations, for three different values dilution factors $d$ = 0.1, 0.5 and 1 of the exit sub-pupils. The basic array parameters are the operating wavelength $\lambda$ = 10 µm, the diameter of the collecting telescopes $D$ = 5 m, their focal length $F$ = 100 m, and the entrance baseline of the array $B$ = 100 m. The combiner parameters $D'$ and $B'$ are constrained via the dilution factor $d$ (see § 3.1). Since our main goal is to comparing the fixed and rotating telescopes options, it is assumed that the brightness of the star and planet are equal, and the main performance metric is a "contrast" factor $C_P(\mathbf{s'})$ defined as:

$$C_P(\mathbf{s'}) = P(\mathbf{s_P}, \mathbf{s'})/S(\mathbf{s'}) , \tag{12}$$

where $S(\mathbf{s'})$ and $P(\mathbf{s_P},\mathbf{s'})$ are the star and planet signals defined by Eqs. 3 and 4 respectively, themselves making use of relations 7–11 in subsection 3.3. In Table 3 are indicated the maximal achievable contrast ratio $C_{Max}$ in the nine considered cases, for both the rotating and fixed telescopes configurations. It also indicates the angular coordinates $u_P$ and $v_P$ of the extrasolar planer, that were determined so as to optimize $C_P(\mathbf{s'})$ in each case. For a fair comparison between both the fixed and rotating array configurations, care was taken that the angles $u_P$ and $v_P$ are equal, i.e. the planets are located at the same position for both options.

In Figure 6 to Figure 8 are presented a series of images obtained for the original Bracewell, DAC and DCB configurations respectively. Each one shows the contrast maps $C_P(\mathbf{s'})$ at different values of the angles $\theta$' and $\theta$' varying from 0° to 90° by steps of 15°, knowing that the full angular range to 360° can be obtained by elementary symmetry transformations. The rotating telescopes configurations are displayed on the left sides of the figures, and the fixed ones on their right sides. In both cases three different dilution factors $d$ = 0.1, 0.5 and 1 are considered. The angular width of each vignette is equal to 4 µrad. The planet locations ($u_P$,$v_P$) are indicated by blue dots. An empirical pseudo-color scale indicates which kind of emitted photons are dominant: blue color for the planet and orange for the star. The contrast curves as function of $\theta$ and $\theta$' for all configurations and dilution factors are plotted on Figure 9 with blue colors for he rotating telescopes configuration and red colors for the fixed telescopes one.

While  to Figure 8 remain essentially illustrative, the contrast curves in Figure 9 reveal strong and rapid variations of the contrast ratio $C_P(\mathbf{s'})$ as function of the angles $\theta$ and $\theta$'. They also show that:

- There is an apparent superiority of the fixed telescopes configuration with respect to the rotating telescopes one, since the maximal contrast values $C_{Max}$ are higher in most cases, at the noticeable exception of the DCB array with $d$ = 0.5. One may also note that excepting the DAC, the higher contrast values are attained with the minimal dilution factor $d$ = 0.1. One also notes that the classical Fizeau configuration where $d$ = 1 seems to be the less efficient.
- As a direct consequence of multi-axial beam combining scheme, a full extinction of the central star is only achievable over limited ranges of the rotation angles $\theta$ and $\theta$'. This leads to defining dedicated observation sequences, which is the scope of the next sub-section.

### 4.3 Observation modes

Fundamentally, nulling interferometers should provide at least two different operating modes that are:

- Mode 1: Exploring and detection mode, searching for unknown extrasolar planets orbiting around their parent star and requiring modulation of the planet signal. Here the signal should oscillate between minimal and a maximal value (typically by a factor of 1/5 or 1/10), always keeping they significantly higher than the star signal (typically > $10^3$).
- Mode 2: Observing mode with long exposure times, staring at an already identified planet in order to analyzing its atmosphere, with no absolute need for signal modulation.

The contrast curves in Figure 10 suggest that both operating modes are only possible over limited ranges of the pupil rotator angle $\theta$', thus requiring an adequate driving sequence. Examples of such observing sequences with 0° ≤ $\theta$' ≤ 90° are illustrated in Figure 10 for three of the previous studied cases. The bounds of the limiting angular ranges are specified in Table 3. Therefore it is concluded that only a discontinuous and limited set of pupil rotator angles $\theta$' is required for modulating the signals emitted by the extrasolar planet, for both the modes 1 and 2. Moreover, these observation sequences may be coupled to a priori knowledge of the planet location, that may be obtained by use of different techniques such as radial velocity, transit observations, or coronagraphy.

Table 2: Planet angular locations ($u_P$,$v_P$) and maximal contrast ratio $C_{Max}$ in the nine considered cases, for both the rotating and fixed telescopes configurations.

| **Array configuration** | | | **Rotating telescopes array** | **Fixed telescopes array** |
|---|---|---|---|---|
| **Two-telescope Bracewell** | $d = 0.1$ | $u_P$ (μrad) | 0.016 | |
| | | $v_P$ (μrad) | 0.394 | |
| | | $C_{Max}$ | 1.2E+02 | 6.3E+03 |
| | $d = 0.5$ | $u_P$ (μrad) | 0.803 | |
| | | $v_P$ (μrad) | 0.016 | |
| | | $C_{Max}$ | 9.0E+00 | 4.6E+03 |
| | $d = 1$ | $u_P$ (μrad) | 0.803 | |
| | | $v_P$ (μrad) | 0.000 | |
| | | $C_{Max}$ | 4.0E+00 | 1.3E+01 |
| **Degenerated Angel Cross (DAC)** | $d = 0.1$ | $u_P$ (μrad) | 0.992 | |
| | | $v_P$ (μrad) | 0.000 | |
| | | $C_{Max}$ | 1.8E+02 | 1.3E+03 |
| | $d = 0.5$ | $u_P$ (μrad) | 1.402 | |
| | | $v_P$ (μrad) | 0.079 | |
| | | $C_{Max}$ | 1.8E+01 | 9.0E+04 |
| | $d = 1$ | $u_P$ (μrad) | 1.701 | |
| | | $v_P$ (μrad) | 0.000 | |
| | | $C_{Max}$ | 1.3E+05 | 1.3E+05 |
| **Linear Dual Chopped Bracewell (DCB)** | $d = 0.1$ | $u_P$ (μrad) | 1.008 | |
| | | $v_P$ (μrad) | 0.000 | |
| | | $C_{Max}$ | 4.3E+05 | > 1.0E+06 |
| | $d = 0.5$ | $u_P$ (μrad) | 0.803 | |
| | | $v_P$ (μrad) | 0.031 | |
| | | $C_{Max}$ | 4.4E+05 | 5.1E+02 |
| | $d = 1$ | $u_P$ (μrad) | 1.701 | |
| | | $v_P$ (μrad) | 0.079 | |
| | | $C_{Max}$ | 1.4E+01 | 3.9E+04 |

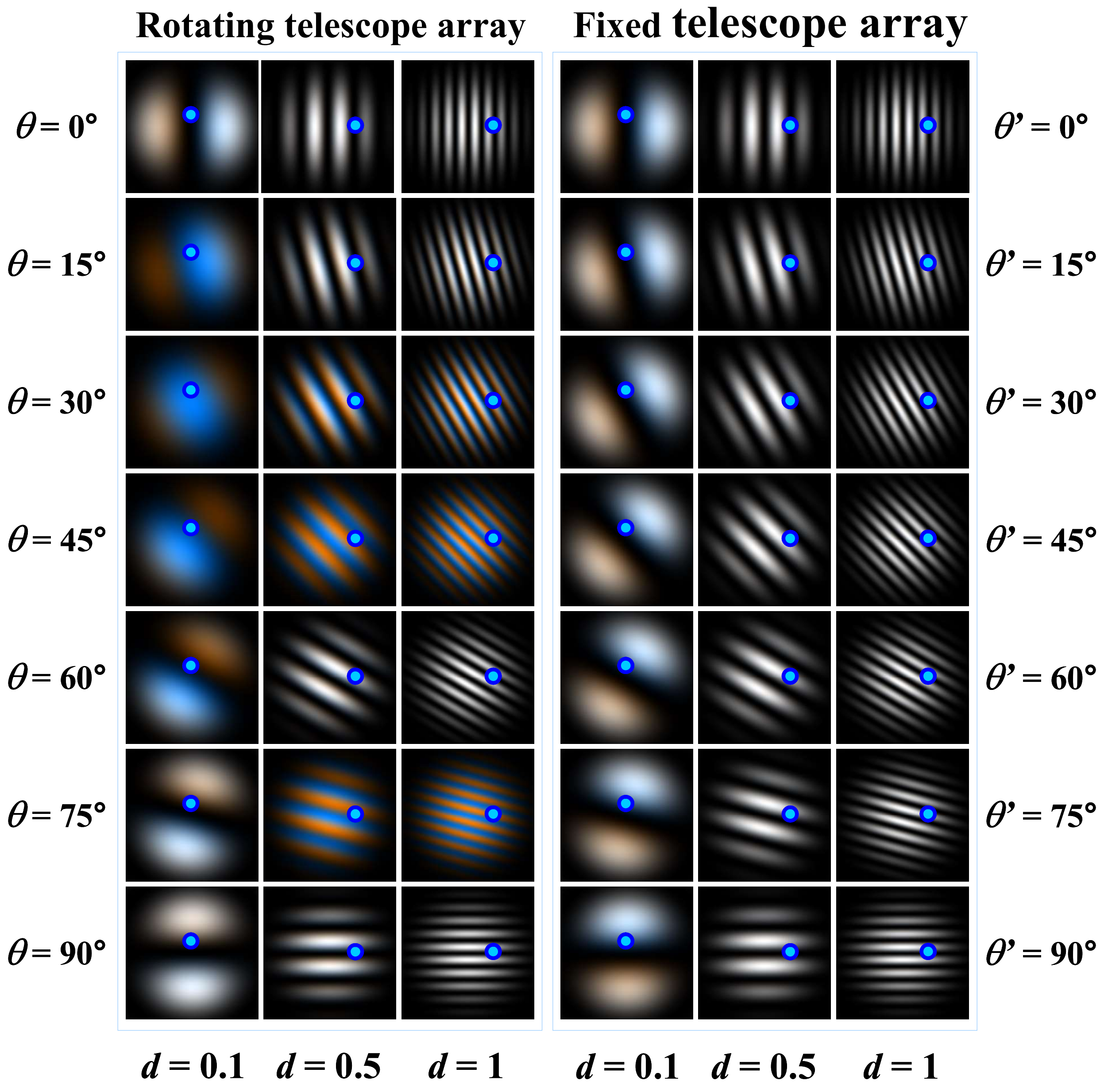

Figure 6: Contrast maps obtained for the original Bracewell configuration at different values of the angles $\theta$' and $\theta$' varying from 0° to 90° by steps of 15° (full angular range can be obtained by symmetry transformations). The rotating telescopes option is displayed on the left side and the fixed telescopes one on the right side. In both cases three different dilution factors $d$ = 0.1, 0.5 and 1 are shown. The angular width of each vignette is equal to 4 µrad. Angular planet locations ($u_P$,$v_P$) are indicated by blue dots. A pseudo-color scale indicates which kind of emitted photons are dominant (blue color for the planet, orange for the star).

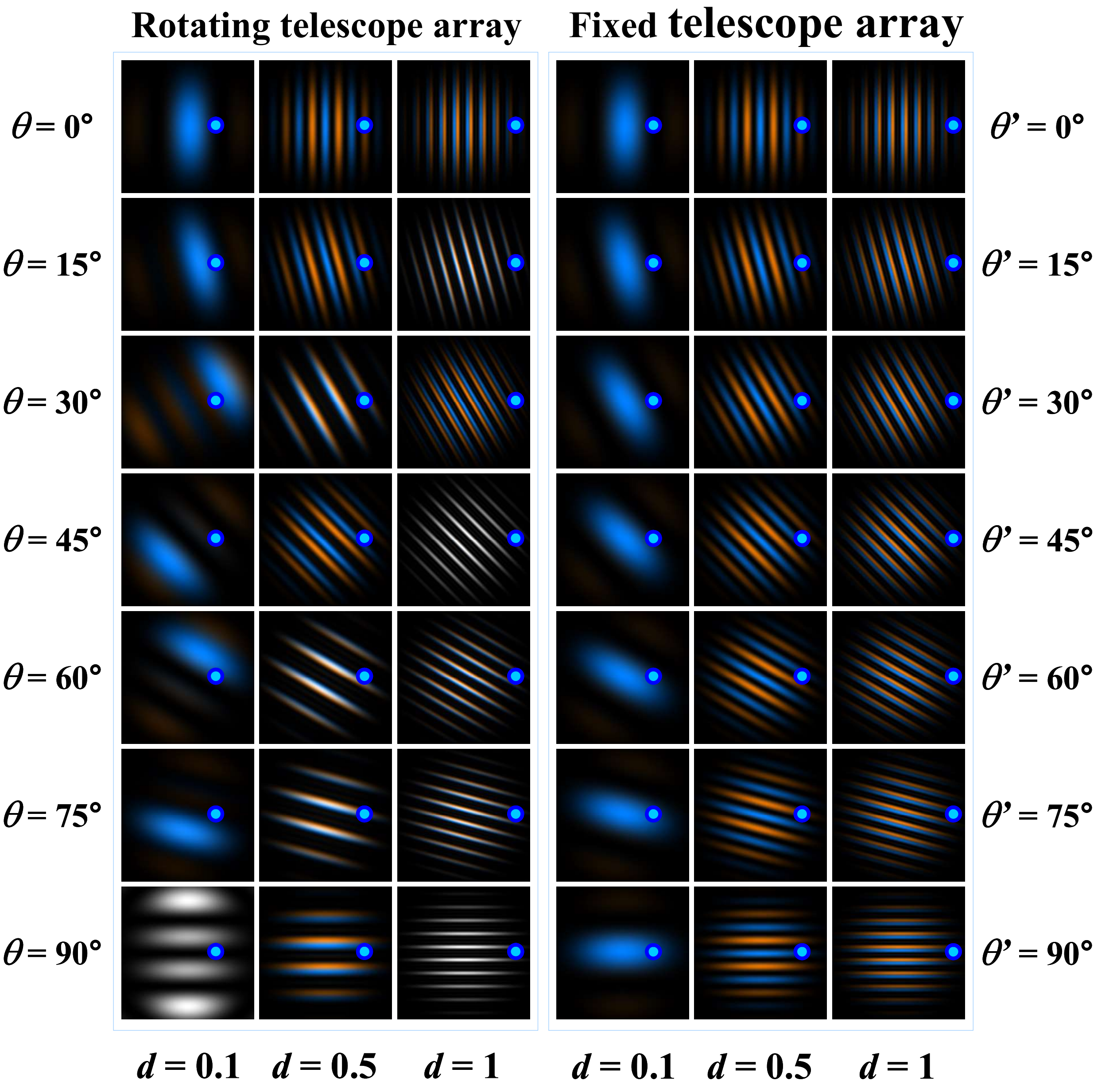

Figure 7: Same illustrations as in  for the DAC configuration.

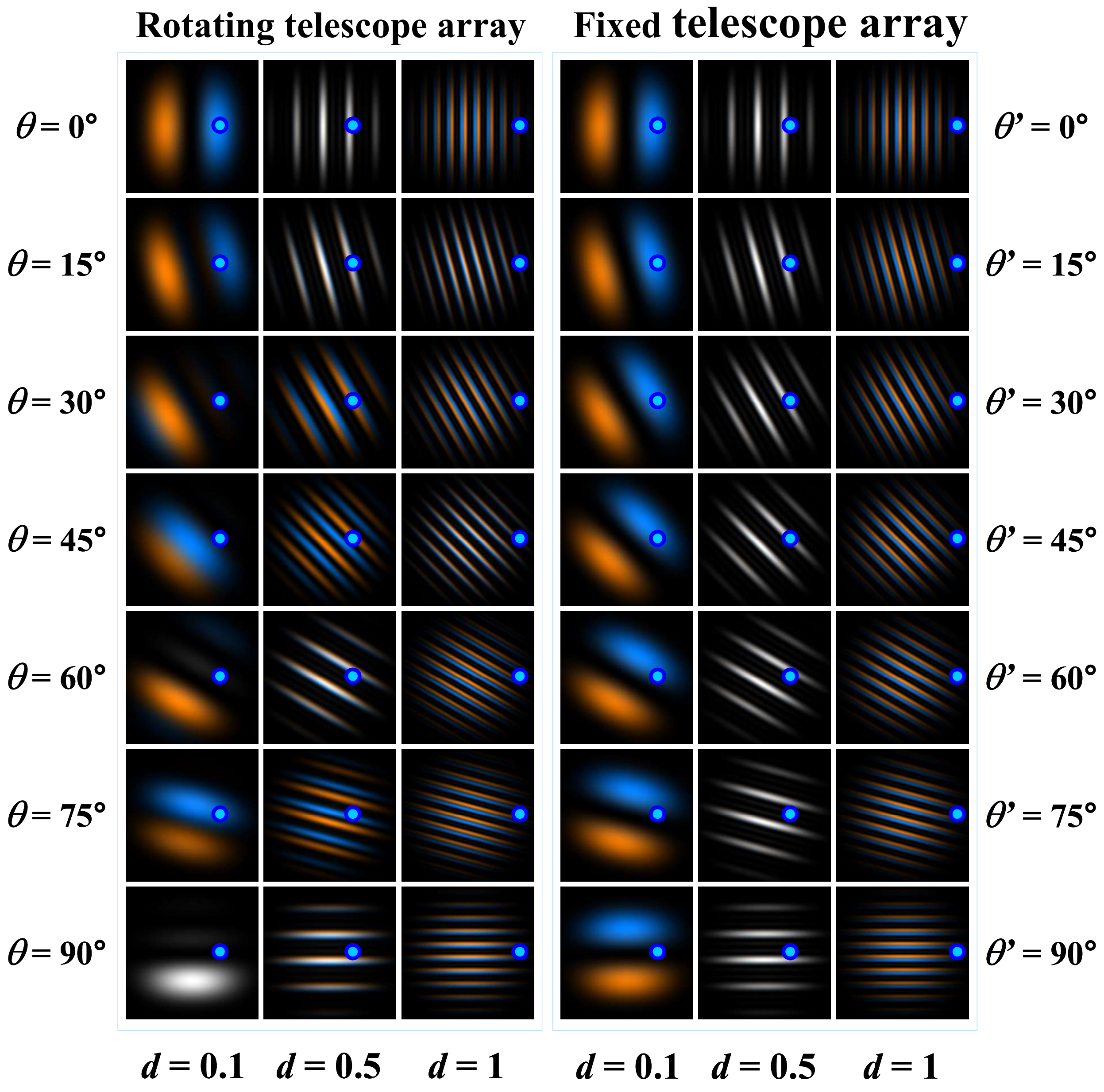

Figure 8: Same illustrations as in  for the DCB configuration.

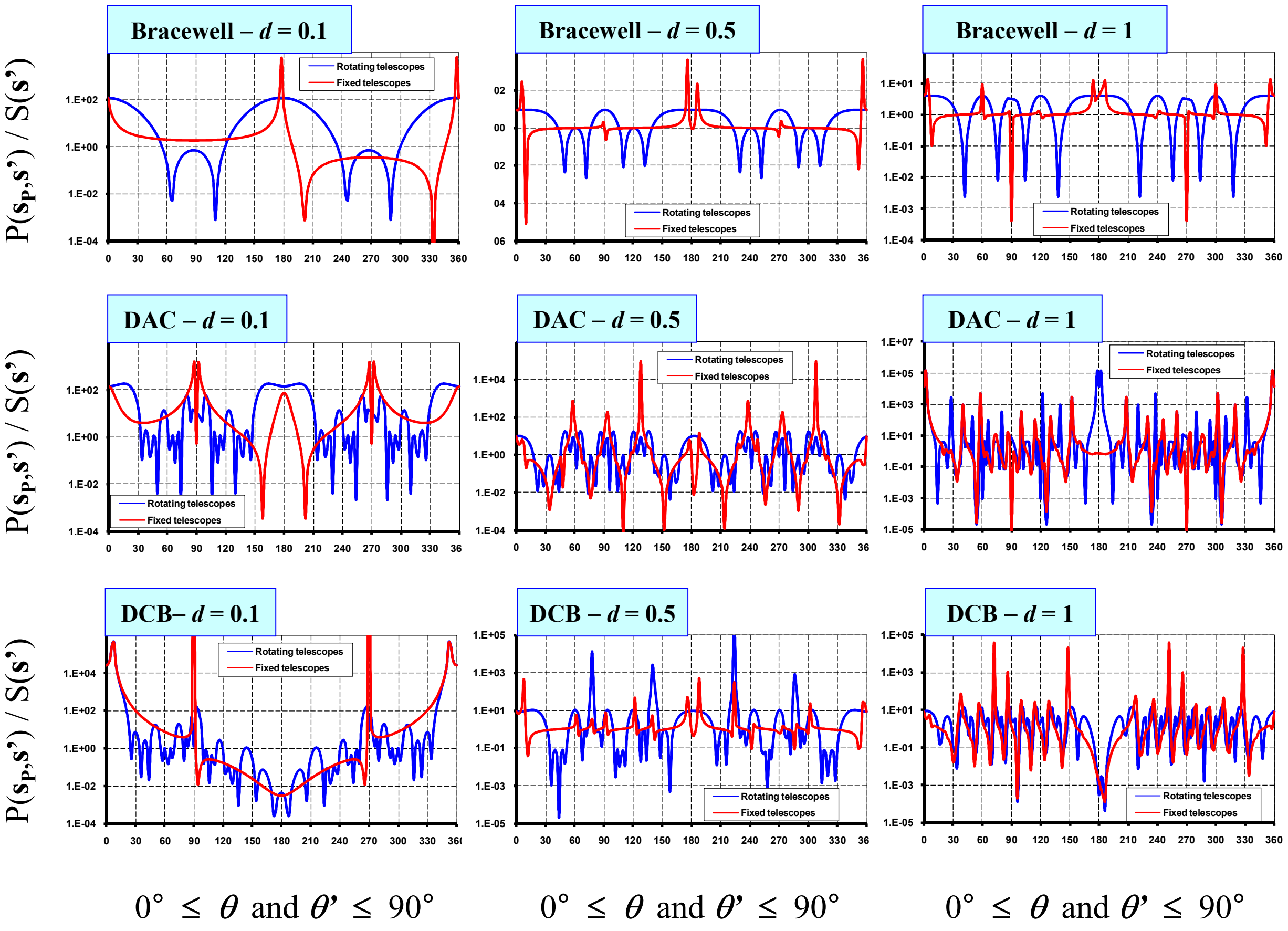

Figure 9: Plots of contrast curves as function of the angles $\theta$ and $\theta$' varying from 0° to 90° by steps of 15° for different configurations and dilution factors (full angular ranges are obtained by symmetry). Top: Original Bracewell configuration. Middle: DAC configuration. Bottom: DCB configuration. The rotating telescopes configuration is shown by the blue curves and the fixed telescopes by red ones.

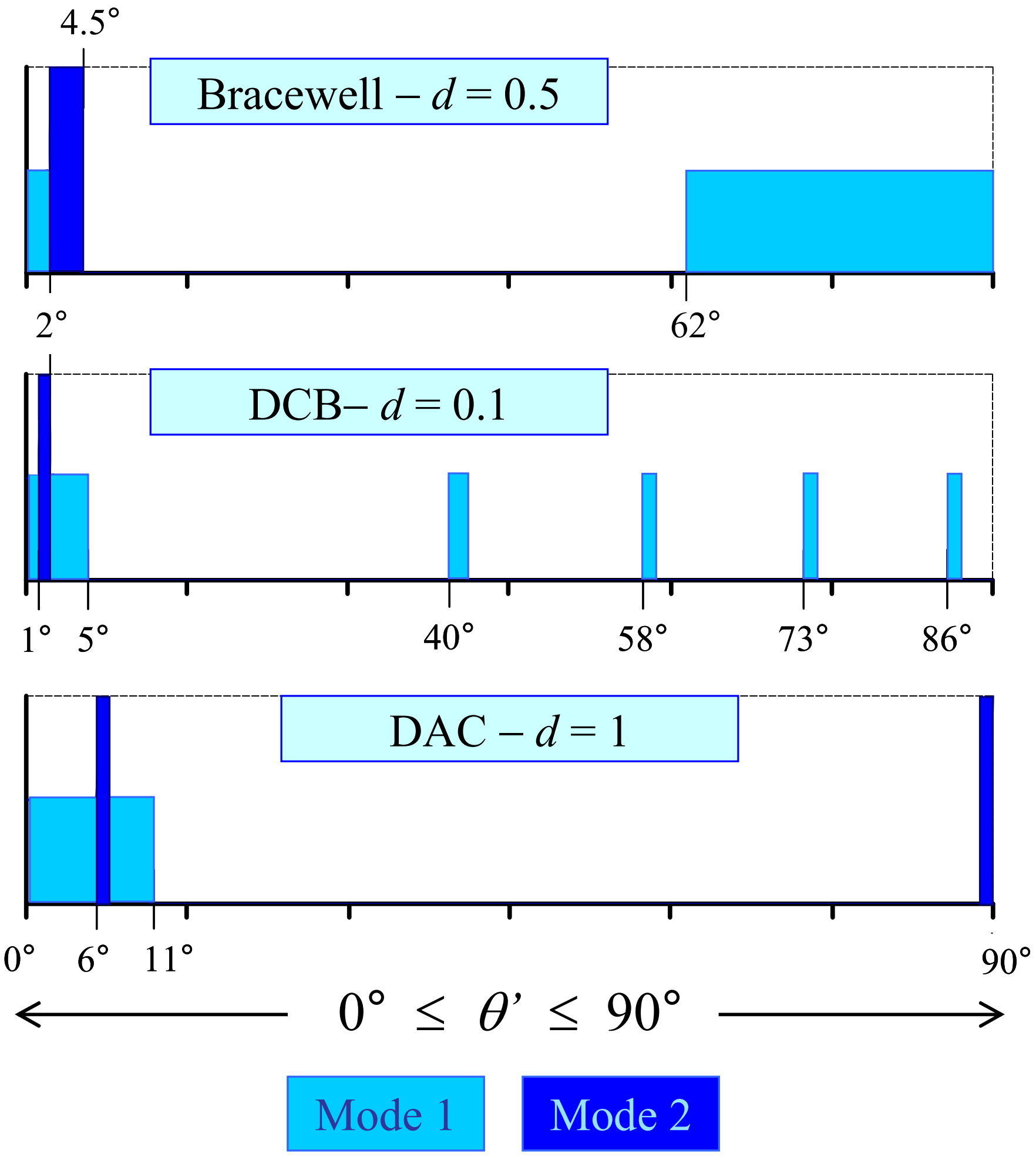

Figure 10: Illustrating different observation sequences with angle $\theta$' varying from 0° to 90° for three different cases (full angular range is obtained by symmetry). Top: Bracewell configuration with $d = 0.5$. Middle: DAC configuration with $d = 1$. Bottom: DCB configuration with $d = 0.1$.

Table 3: Examples of observing sequences for modes n°1 and 2 in the range $\theta' \in [0°,90°]$.

| **Configuration / dilution factor** | **Modes 1 and 2, maximal planet signal** | **Mode 1, lower planet signal** |
|---|---|---|
| Bracewell with $d = 0.5$ | $\theta' \in$ [2.2°, 4.8°] | $\theta' \in$ [0°, 2.2°]; [61.7° ,90°] |
| DAC with $d = 1$ | $\theta' \in$ [1.2°, 2.3°] | $\theta' \in$ [0°, 1.2°]; [2.3°, 5.2°]; [38.7°, 40.8°]; [57.7°, 58.3°]; [72.7°, 73.3°]; [86.2°, 86.8°] |
| DCB with $d = 0.1$ | $\theta' \in$ [6.7°, 7.3°]; [89.7°, 90°] | $\theta' \in$ [0°, 6.7°]; [7.3°, 11.3°] |

# 5 CONCLUSION

In this paper was described the conceptual design of a nulling interferometer operating in space without needing a rotating telescope array. It is essentially composed of:

- A set of two ore more fixed collecting telescopes arranged in a linear array, either onboard of free-flying spacecrafts or mounted onto a truss structure,
- Altazimuthal drives of the collecting telescopes, allowing them to exploring a wide field of view,
- Long range delay lines systems integrated into the spacecrafts,
- Multi-axial beam combining optics comprising a pupil rotator made of prisms or mirrors, and located in the central spacecraft,
- A dedicated observation sequence allowing both detection and characterization modes of extrasolar planets.

This versatile concept is compatible with any kind of achromatic phase-shifter and WFE filtering device, which are two vital components of a nulling interferometer. Numerical simulations demonstrated that the achievable performance in terms of contrast ratio is equivalent or even higher than that of a classical rotating telescopes array. The major drawback of the fixed telescopes concept resides in the loss of information about one of the two angular coordinates of the observed extrasolar planet. On the other hand, a huge economy in fuel consumption shall result and allow for longer mission durations. Also, it appreciably reduces the risks of loosing one or more spacecrafts and aborting the mission before its end, thus facilitating the development of space borne nulling interferometry technique in the forthcoming years. Finally, it should be noted that the fixed telescopes concept is also applicable to a fibered nulling interferometer, with additional ability to adjusting or modulating the baseline *B*' of the exit sub-pupils.